\documentclass[pre,twocolumn,showpacs,preprintnumbers,floatfix]{revtex4}

\usepackage{makeidx}
\usepackage{amssymb}
\usepackage{graphicx}

\begin{document}

\title{\textbf{Fluctuating dynamics at the quasiperiodic onset of chaos,
Tsallis }$q$\textbf{-statistics and Mori's }$q$\textbf{-phase thermodynamics}%
}
\author{H. Hern\'{a}ndez-Salda\~{n}a}
\email{hugo@fisica.unam.mx}
\author{ A. Robledo}
\email{robledo@fisica.unam.mx }
\affiliation{Instituto de F\'{\i}sica,\\
Universidad Nacional Aut\'{o}noma de M\'{e}xico,\\
Apartado Postal 20-364, M\'{e}xico 01000 D.F., Mexico.}
\date{.}

\begin{abstract}
We analyze the fluctuating dynamics at the golden-mean transition to chaos
in the critical circle map and find that trajectories within the critical
attractor consist of infinite sets of power laws mixed together. We
elucidate this structure assisted by known renormalization group (RG)
results. Next we proceed to weigh the new findings against Tsallis' entropic
and Mori's thermodynamic theoretical schemes and observe behavior to a large
extent richer than previously reported. We find that the sensitivity to
initial conditions $\xi _{t}$ has the form of families of intertwined $q$%
-exponentials, of which we determine the $q$-indexes and the generalized
Lyapunov coefficient spectra $\lambda _{q}$. Further, the dynamics within
the critical attractor is found to consist of not one but a collection of
Mori's $q$-phase transitions with a hierarchical structure. The value of
Mori's `thermodynamic field' variable $\mathsf{q}$ at each transition
corresponds to the same special value for the entropic index $q$. We discuss
the relationship between the two formalisms and indicate the usefulness of
the methods involved to determine the universal trajectory scaling function $%
\sigma $ and/or the ocurrence and characterization of dynamical phase
transitions.
\end{abstract}

\maketitle
\preprint{123XXXX}

\section{Introduction}

The unfamiliar dynamics at critical attractors of even the standard one
dimensional nonlinear iterated maps \cite{schuster1}-\cite{hilborn1} is of
current interest to statistical physicists; it provides insights into
properties of systems at which the key supports of the canonical theory,
phase space mixing and ergodicity, break down. At these attractors the
indicators of chaos withdraw, such as the fast rate of separation of
initially close-by trajectories. There is a small number of known routes to
chaos and it is of interest to know if at their thresholds the dynamics
displays similar features and if so to what extent. If present, these
similarities would apply above the known universality of each type of
critical attractor.

Critical attractors of nonlinear low-dimensional maps have always drawn much
interest because they demonstrate the ways by which nonlinear systems bridge
periodic and chaotic motion \cite{schuster1}-\cite{hilborn1}. The
geometrical properties of these attractors have been known for a long time 
\cite{schuster1}-\cite{hilborn1}, but their dynamical properties are even
now perplexing and still need to be fully understood. Renewed interest in
this subject \cite{tsallis2} - \cite{robledo8} has focused on the
applicability of $q$-statistics, the generalization of the Boltzmann-Gibbs
(BG) statistical mechanics proposed by Tsallis \cite{tsallis0}, \cite%
{tsallis1}, to these nonlinear systems. Important examples of critical
attractors are the onset of chaos via period doubling, intermittence and
quasiperiodicity, the three universal routes to chaos in dissipative maps
displayed by the prototypical logistic and circle maps \cite{schuster1} - 
\cite{hilborn1}. Perhaps the most interesting types of critical attractors
are critical strange nonchaotic attractors \cite{grebogi1} as in the period
doubling and quasiperiodic onset of chaos. These are geometrically
complicated multifractal attractors with a vanishing Lyapunov coefficient $%
\lambda _{1}$ and a sensitivity to initial conditions $\xi _{t}$ that does
not converge to any single-valued function but instead displays a
fluctuating pattern that grows as a power law in time $t$ \cite{grassberger1}
- \cite{mori1}.

Trajectories \textit{within} a critical attractor show self-similar temporal
structures, they preserve memory of their previous locations and do not have
the mixing property of truly chaotic trajectories \cite{mori1}. A special
version of the 'thermodynamic formalism' was adapted long ago \cite{politi1}-%
\cite{mori1} to study the dynamics at critical attractors and quantitative
results were obtained that provided a first understanding, particularly
about the envelope of the fluctuating $\xi _{t}$ and the ocurrence of a
so-called `$q$-phase' dynamical phase transition \cite{mori3} - \cite{mori1}%
. As shown below we uncover a complete and rich dynamical structure for $\xi
_{t}$ and clarify its relationship to families of $q$-phase transitions. We
explain how these transitions relate to distinct regions of the multifractal
attractor, and their manifestation in the universal jump discontinuities of
the trajectory scaling function $\sigma $ \cite{feigenbaum1} - \cite%
{mainieri1}. We clarify also the relationship of the thermodynamic approach
with $q$-statistics with regards to the dynamical properties studied.

Recently, rigorous results have been developed \cite{robledo1}-\cite%
{robledo8} that support the validity of $q$-statistics (outlined in the next
paragraph) for the critical attractors associated with the intermittency and
period-doubling routes to chaos, i.e. the tangent bifurcation and the
accumulation point of the pitchfork bifurcations, but studies of the same
type have not been carried out for the quasiperiodic route to chaos. Here we
present results on this route by considering the proverbial golden-mean
onset of chaos in the critical circle map \cite{schuster1}, \cite{hilborn1}
and show that there are elements of universality in their dynamical
properties as they appear in close correspondence with those recently
analyzed for the Feigenbaum attractor \cite{robledo8}.

The central point of the $q$-statistics with regards to the description of
the dynamics of critical attractors is a sensitivity to initial conditions $%
\xi _{t}$ related to the $q$-exponential functional form, i.e. the `$q$%
-deformed' exponential function $\exp _{q}(x)\equiv \lbrack
1-(q-1)x]^{-1/(q-1)}$ \cite{tsallis2}, \cite{tsallis3}. From such $\xi _{t}$
one or more spectra of $q$-generalized Lyapunov coefficients $\lambda _{q}$
can be determined \cite{robledo2}, \cite{robledo8}. The $\lambda _{q}$ are
dependent on the initial position and each spectrum can be examined by
varying this position. The $\lambda _{q}$ satisfy an identity $\lambda
_{q}=K_{q}$ where $K_{q}$ is an entropy production rate based on the Tsallis
entropy $S_{q}$, defined in terms of the $q$-logarithmic function $\ln
_{q}y\equiv (y^{1-q}-1)/(1-q)$, the inverse of $\exp _{q}(x)$ \cite{tsallis2}%
, \cite{robledo2}, \cite{robledo8}.

Here we find that $\xi _{t}$ at the golden-mean onset of chaos takes the
form of a family of interwoven $q$-exponentials. The $q$-indexes appear in
conjugate pairs, $q$ and $Q=2-q$, as these correspond to switching starting
and finishing trajectory positions. We show that $q$ and $Q$ are related to
the occurrence of pairs of dynamical $q$-phase transitions that connect
qualitatively different regions of the attractor \cite{mori3} - \cite{mori1}%
. These transitions are identified as the source of the special values for
the entropic index $q$, and these values are determined by the universality
class parameters to which the attractor belongs. In our present case the
parameters are the golden-mean winding number $w_{gm}=(\sqrt{5}-1)/2\simeq
0.618034$ and the universal constants (in the sense of Feigenbaum's
trajectory scaling function $\sigma $ \cite{feigenbaum1} - \cite{mainieri1})
that measure the power-law clustering of iterate positions. The most
prominent of the universal constants is $\alpha _{gm}\simeq -1.288575$, that
appears in the description of the most rarefied region and (as $\alpha
_{gm}^{3}$) of the most crowded region of the multifractal attractor.

We are interested here in determining the detailed dependence of the
aforementioned dynamical structure for the golden-mean chaos threshold on 
\textit{both} the initial position and the observation time $t$ as this
dependence is preserved by the infinite memory of trajectories. To this
purpose we recall in the following Section 2 the familiar features of the
circle map, and the properties of the renormalization group (RG) fixed-point
map for a cubic inflection point with zero slope. Then in Section 3 we
describe the organization of the superstable trajectory at the chaos
threshold into families of positions that lie along power laws in time.
Assisted by these properties, in Section 4 we determine the sensitivity $\xi
_{t}$ for trajectories that have their starting and finishing positions at
the most crowded and most sparse regions of the multifractal attractor. In
Section 5 we explain how the properties for $\xi _{t}$ and its $\lambda _{q}$
spectrum are related to the occurrence of dynamical $q$-phase transitions
and corroborate previously obtained numerical results. In Section 6 we
express $\xi _{t}$ in terms of the discontinuities of the trajectory scaling
function $\sigma $ that measures local contraction rates within the
multifractal attractor. From this relationship we obtain results for
trajectories that involve other multifractal regions and show that the
dynamics consists of a hierarchy of $q$-indexes and $q$-phase transitions.
In Section 7 we summarize our results and further disscuss the links between
the thermodynamic and $q$-statistical schemes and call attention to the
usefulness of the latter.

\section{The golden-mean onset of chaos}

To assist our presentation in the following sections we recall here basic
features of the circle map \cite{schuster1}, \cite{hilborn1}. The map is
given by%
\begin{equation}
f_{\Omega ,K}(\theta )=\theta +\Omega -K/2\pi \sin 2\pi \theta ,\;\text{mod}%
\;1,  \label{circle1}
\end{equation}%
where the control parameters $\Omega $\ and $K$ are the bare winding number
and the amount of nonlinearity, respectively. An important parameter for the
description of the structure of the trajectories of this map is the dressed
winding number%
\begin{equation}
w\equiv \lim_{t\rightarrow \infty }\frac{\theta (t)-\theta (0)}{t},
\label{dressed1}
\end{equation}%
the average increment of $\theta (t)$ per iteration. The map is monotonic
and invertible for $K<1$, develops a cubic inflexion point at $\theta =0$
(or $\theta =1$, see Figs. 1) for $K=1$, and becomes nonmonotonic and
noninvertible for $K>1$. For $K<1$ trajectories are periodic (locked motion)
when $w$ is rational and quasiperiodic (unlocked motion) when $w$ is
irrational. The winding number $w(\Omega )$ forms a `devil's staircase'
making a step at each rational $p/q$ and remaining constant $w=p/q$ for a
range of $\Omega $. For $K=1$, the critical circle map, locked motion covers
the entire $\Omega $ interval leaving only a multifractal set of unlocked $%
\Omega $. The threshold to chaos is obtained at the accumulation points of
sequences of control parameter values of families of periodic trajectories.
At these accumulation points the trajectories are quasiperiodic. For $K>1$
the regions of periodic motion (Arnold tongues) overlap leading to chaotic
motion.

\begin{figure}[tbp]
\includegraphics[width=0.9\columnwidth,angle=-90,scale=0.7]{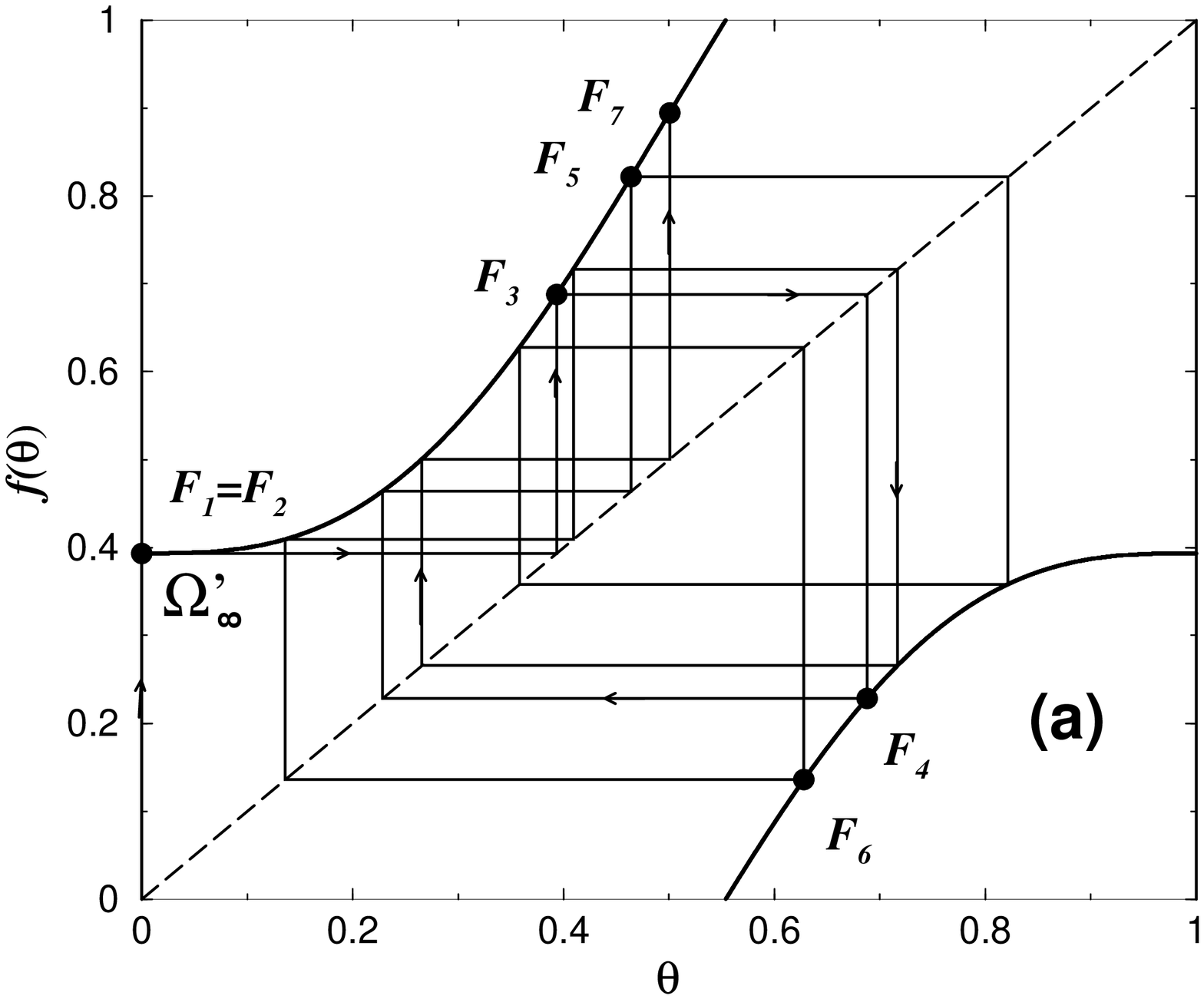} %
\includegraphics[width=0.9\columnwidth,angle=-90,scale=0.7]{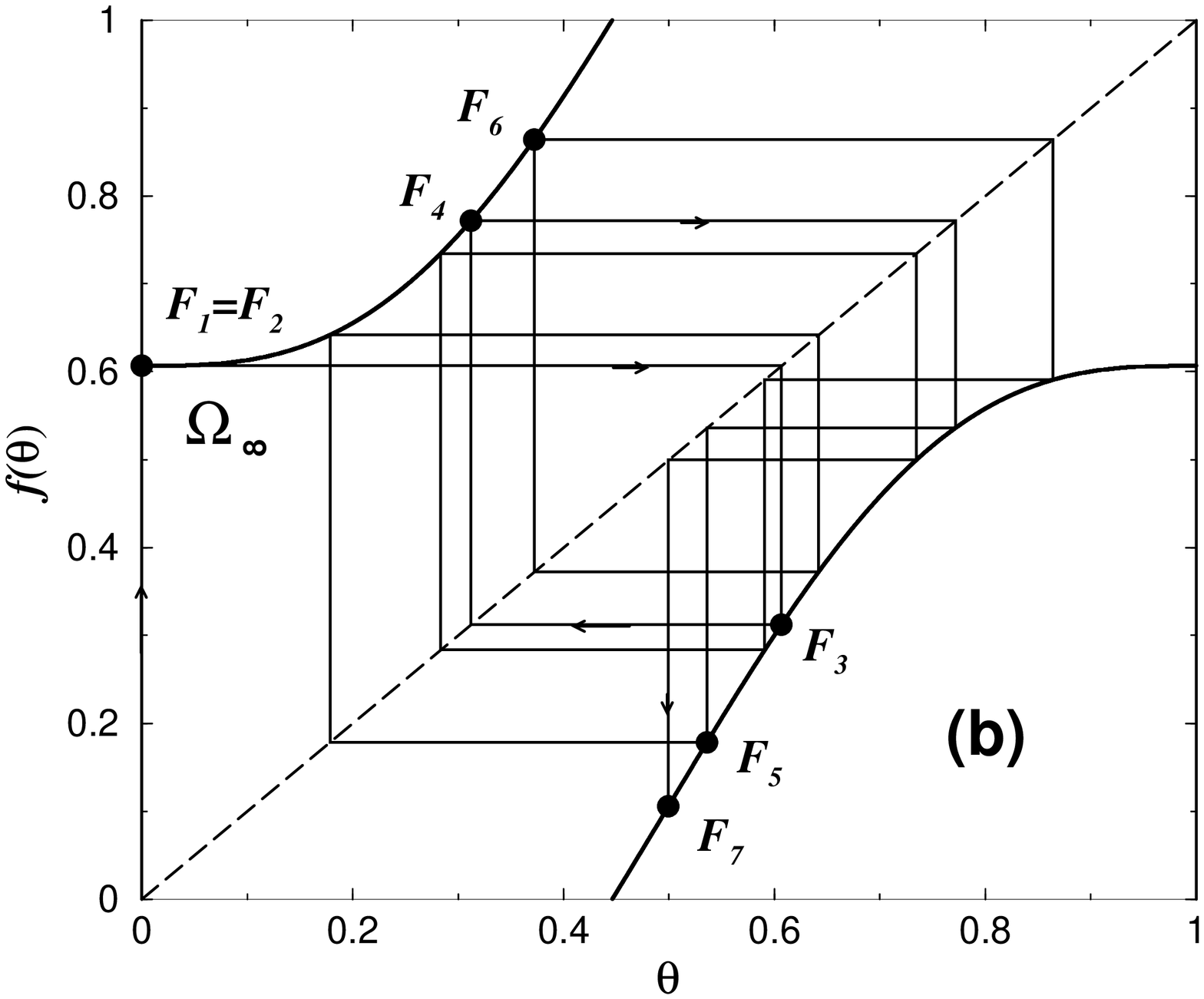} 
\caption{{\protect\small Critical circle map }$K=1${\protect\small \ and
supercycle trajectories. a) }$\Omega =\Omega _{\infty }^{\prime }$%
{\protect\small . b) }$\Omega =\Omega _{\infty }${\protect\small . The
labels }$F_{n}${\protect\small \ correspond to trajectory times given by
Fibonacci numbers.} }
\end{figure}

A standard way to study the quasiperiodic route to chaos is to fix the value
of $K=1$ and select an irrational $w$. Then a choice is made of successive
rational values for $\Omega $ to obtain an infinite sequence of winding
numbers that approach $w$. A convenient sequence of such rational values for 
$\Omega $ are given by consecutive truncations of the continued fraction
expansion of $w$. A well-studied, an perhaps the most interesting, case is
the sequence of rational approximants to $w_{gm}=(\sqrt{5}-1)/2\simeq
0.618034$, the reciprocal of the golden mean, that yields the winding
numbers $w_{n}=F_{n-1}/F_{n}$, where $F_{n}$ are the Fibonacci numbers $%
F_{n+1}=$ $F_{n}+F_{n-1}$. Within the range of $\Omega $ for which $w=$ $%
F_{n-1}/F_{n}$ one observes trajectories of period $F_{n}$, and therefore a
route to chaos consists of an infinite family of periodic orbits with
increasing periods of values $F_{n}$, $n\rightarrow \infty $. It has proved
useful to single out from these families the periodic trajectories with the
`superstable' property. We recall that a superstable orbit \cite{schuster1}, 
\cite{hilborn1} is an orbit of period $T$ such that $df^{T}(\theta
_{0})/d\theta =0$. In our case a superstable orbit is one that contains as
one of its positions $\theta =0$. Since the derivative of $f_{\Omega
_{n},1}(\theta )$ at $\theta =0$ vanishes the Lyapunov coefficient $\lambda
_{1}$ diverges to $-\infty $ for these family of orbits. The values of $%
\Omega $, $\Omega _{n}$, $n=1,2,...$, at which they occur can be determined
from their definition when written as $f_{\Omega _{n},1}^{F_{n+1}}(0)=F_{n}$%
, $n=1,2,...$The solutions of this equation lie on zigzag paths approaching
the limit $\Omega _{\infty }\simeq 0.606661$ which corresponds to $w_{gm}$.
As the golden mean $w_{gm}$ arises from the condition $w=1-w^{2}$, a second
superstable family of trajectories is determined by considering the winding
numbers $w_{n}^{\prime }=F_{n-2}/F_{n}$, with $\Omega _{n}^{\prime }$, $%
n=1,2,...$, obtained in turn from $f_{\Omega _{n}^{\prime
},1}^{F_{n+1}}(0)=F_{n-1}$. The $\Omega _{n}^{\prime }$ converge to $\Omega
_{\infty }^{\prime }=1-\Omega _{\infty }\simeq 0.393339$ which corresponds
to $w_{gm}^{2}\simeq \allowbreak 0.381\,966$. In Figs. 1a and 1b we show the
critical map for $\Omega =\Omega _{\infty }^{\prime }$ and $\Omega =\Omega
_{\infty }$, respectively, together with a portion of the superstable
trajectory at the chaos threshold. In Fig. 2 we show the values for the
Lyapunov coefficients $\lambda _{1}$ for $K=1$ as a function of $\Omega $
where the locations of the two superstable families are clearly seen and
those for their accumulation points are indicated.

\begin{figure}[tbp]
\includegraphics[width=0.9\columnwidth,angle=-90,scale=0.8]{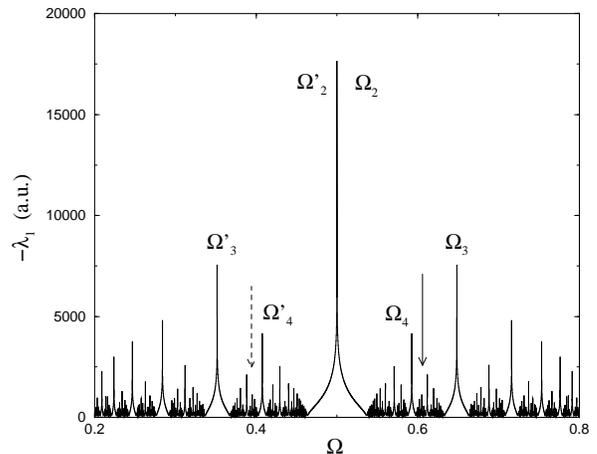} 
\caption{{\protect\small The Lyapunov coefficient }$\protect\lambda _{1}$%
{\protect\small \ for the critical circle map }$K=1${\protect\small \ as a
function of winding number. The supercycle divergences to }$-\infty $%
{\protect\small \ can be clearly appreciated and the locations for the two
accumulation points at }$\Omega _{\infty }^{\prime }${\protect\small \ and }$%
\Omega _{\infty }${\protect\small \ are indicated by the arrows. } }
\end{figure}

A celebrated development \cite{schuster1}, \cite{hilborn1} dating from the
early 80's is the recognition that the quasiperiodic route to chaos displays
universal scaling properties. The analysis was based on an RG approach
analogous to that for the period doubling cascade, and can be suitably
discussed in terms of the critical circle map. The universal properties are
obtained from the fixed-point map $g(\theta )$ of an RG transformation that
consists of functional composition and rescaling appropriate for maps with a
zero-slope cubic inflection point. The fixed-point map satisfies 
\begin{equation}
g(\theta )=\alpha _{gm}g(\alpha _{gm}g(\theta /\alpha _{gm}^{2}),
\label{fixed-point1}
\end{equation}%
where $\alpha _{gm}\simeq -1.288575$ is a universal constant \cite{schuster1}%
, \cite{hilborn1}. This constant describes the scaling of the distance $%
d_{n} $ from $\theta =0$ to the nearest element of the orbit with $w_{n}$,
i.e. $d_{n}=$ $f_{\Omega _{n},1}^{F_{n}}(0)-F_{n-1}$. The scaling is given
by $\lim_{n\rightarrow \infty }(d_{n}/d_{n+1})=\alpha _{gm}$. Related to the
chaos threshold at the golden mean $w_{gm}$ there is a multifractal set (the
attractor) that measures the different degrees of clustering of iterate
positions. This multifractal has a dimension spectrum $f(\alpha )$ within a
range of values between $\alpha _{\min }$ and $\alpha _{\max }$ \cite%
{jklps85}. The most rarefied region of the multifractal corresponds to $%
\alpha _{\max }=-\ln w_{gm}/\ln \left\vert \alpha _{gm}\right\vert \simeq
1.897995$, and this region is mapped onto the multifractal's most crowded
region characterized by $\alpha _{\min }=-\ln w_{gm}/3\ln \left\vert \alpha
_{gm}\right\vert \simeq 0.632665$ \cite{jklps85}. The chaos threshold for
especial values of the winding number is known as local universality \cite%
{schuster1}.

\section{Structure of trajectories within the attractor}

We consider the last of the superstable trajectories, with infinite period
when $\Omega =\Omega _{\infty }$ or $\Omega _{\infty }^{\prime }$ and
starting at $\theta (0)=0$. As we shall see all other trajectories within
the attractor can be obtained from it. We find it convenient to plot the
positions $\theta (t)$ and their iteration times $t$ in logarithmic scales
as shown in Figs. 3a and 3b, obtained for $\Omega _{\infty }$ and $\Omega
_{\infty }^{\prime }$, respectively, and where the labels record the times
at which positions are reached. A distinctive feature in these figures is
that positions fall along straight diagonal lines, an indication of embedded
power law behavior. Notice that the positions of the main diagonal in Fig.
3a correspond to the times $F_{2n}$, $n=1,2,3,...$ The succeeding diagonals
above it appear grouped together. Above the main diagonal there are $k=2$
close-by diagonals the first with positions that correspond to times of the
form $2F_{2n}$ and the second for times of the form $F_{2n}+F_{2n-2}$. Above
these two diagonals there is a group of $k=3$ diagonals and above them a
group of $k=4$, etc. The times for the positions along each diagonal in the
group of $k$ diagonals are, from the bottom to the top, of the form $kF_{2n}$%
, $(k-1)F_{2n}+F_{2n-2}$, $(k-2)F_{2n}+2F_{2n-2}$,..., $F_{2n}+(k-1)F_{2n-2}$%
, $n=1,2,3,...$ The same structure is observed in Fig. 3b with $F_{2n}$
replaced by $F_{2n+1}$. The slope shared by all the diagonal straight lines
in Figs. 3a and 3b is $-\ln \left\vert \alpha _{gm}\right\vert /\ln w_{gm}$.
Note too that the positions for $F_{2n}$ in Fig. 3a appear along the top of
Fig. 3b and approach asymptotically the horizontal line $\theta =1$. And
vice versa, the positions for $F_{2n+1}$ in Fig. 3b appear along the top of
Fig. 3a and approach asymptotically the horizontal line $\theta =1$.
Analogous features hold for the positions of the groups of subsequences with 
$k>1$. A similar pattern of time subsequence positions arranged along
parallel decreasing straight lines is observed in the plot of $\ln
\left\vert \Omega _{\infty }-\theta (t)\right\vert $ vs $\ln t$ (or
identically, $\ln \left\vert \Omega _{\infty }^{\prime }-\theta
(t)\right\vert $ vs $\ln t$) that is shown in Fig. 4. As it can be observed
there, the power-law diagonals along which positions fall appear again
arranged in groups, each with an increasing number of members $k=1,2,3,..$.
The times for the positions along the group of $k$ diagonals are, from the
bottom to the top, of the form $kF_{n}+1$, $(k-1)F_{n}+F_{n-2}+1$, $%
(k-2)F_{n}+2F_{n-2}+1$,..., $F_{n}+(k-1)F_{n-2}+1$, $n=1,2,3,...$, and the
slope shared by all the diagonals is $-3\ln \left\vert \alpha
_{gm}\right\vert /\ln w_{gm}$.

\begin{figure}[tbp]
\includegraphics[width=.9\columnwidth,angle=-90,scale=0.55]{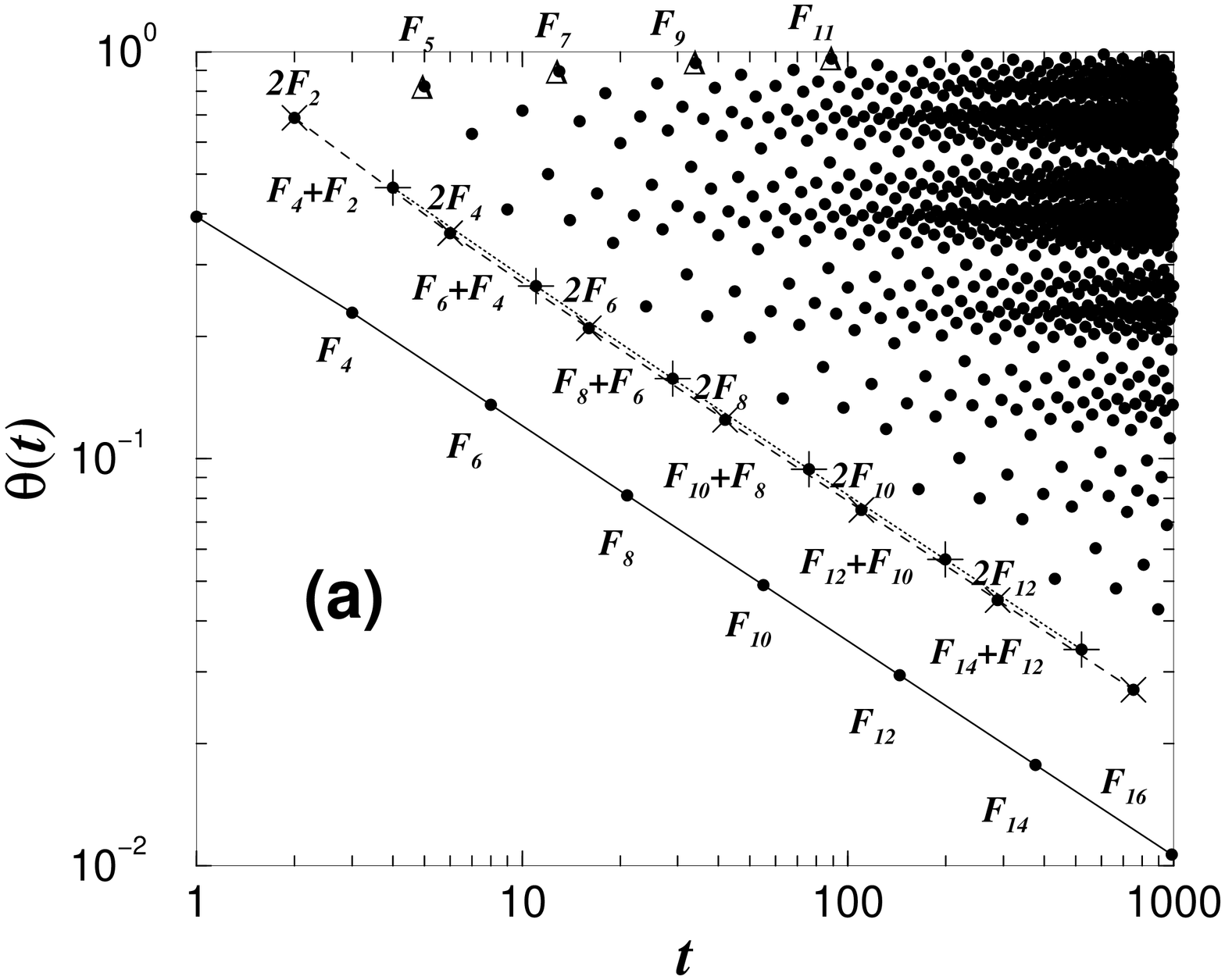} %
\includegraphics[width=.9\columnwidth,angle=-90,scale=0.55]{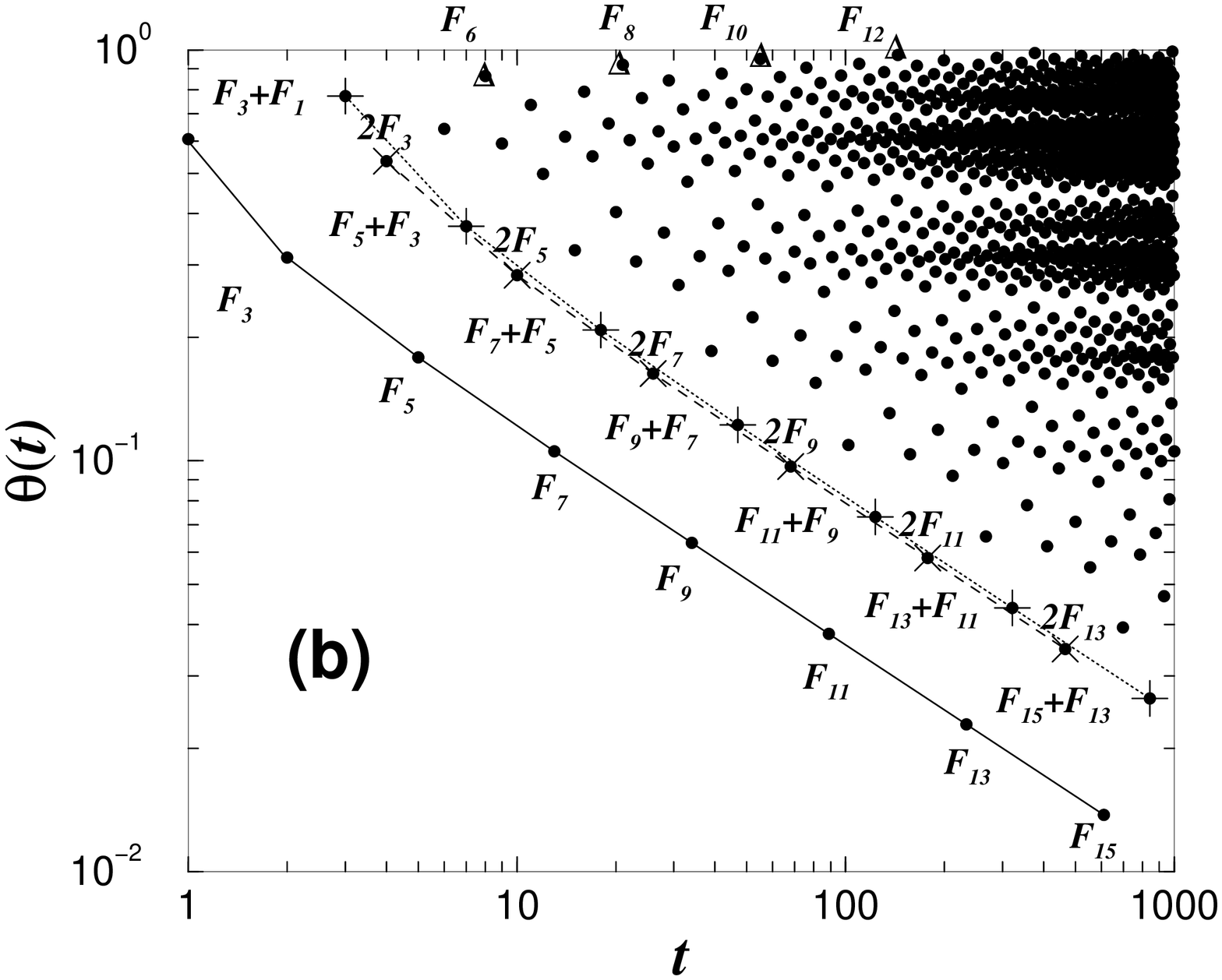} 
\caption{ {\protect\small a) Positions }$\protect\theta (t)${\protect\small %
\ vs }$t${\protect\small \ in logarithmic scales for the orbit with initial
condition }$\protect\theta (0)=0${\protect\small \ at }$\Omega _{\infty
}^{\prime }${\protect\small \ of the critical circle map }$K=1$%
{\protect\small . The lines are guides to the eye. The labels indicate
iteration time }$t${\protect\small , the dashed line and symbols }$\times $%
{\protect\small \ correspond to times of the form }$2F_{2n}${\protect\small %
, and the dotted line and symbols }$+${\protect\small \ correspond to times
of the form }$F_{2n}+F_{2n-2}${\protect\small . The symbols }$\Delta $%
{\protect\small \ correspond to the main diagonal positions at }$F_{2n+1}$%
{\protect\small \ in (b). b) Same as a) with }$\Omega _{\infty }^{\prime }$%
{\protect\small \ replaced by }$\Omega _{\infty }${\protect\small . The
dashed line and symbols }$\times ${\protect\small \ correspond to times of
the form }$2F_{2n+1}${\protect\small , and the dotted line and symbols }$+$%
{\protect\small \ correspond to times of the form }$F_{2n+1}+F_{2n-1}$%
{\protect\small . The symbols }$\Delta ${\protect\small \ correspond to the
main diagonal positions at }$F_{2n}${\protect\small \ in (a). See text for
description.} }
\end{figure}

More specifically, the trajectory starting at $\theta (0)=0$ maps out the
attractor in such a way that the values of succeeding (time-shifted $\tau
=t+1$) positions $\theta _{\tau }$ form subsequences with a common power-law
decay of the form $\tau ^{-1/1-Q}$ with $Q=1-\ln w_{gm}/\ln \left\vert
\alpha _{gm}\right\vert \simeq 2.897995$. (To avoid confusion we denote as $%
\theta (t)$ position at time $t$ and as $\theta _{\tau }$ the same position
at shifted time $\tau $). That is, the \textit{entire} attractor can be
decomposed into position subsequences generated by the time subsequences $%
\tau =(k-l)F_{n}+lF_{n-2}$, each obtained by running over $n=1,2,3,...$ for
fixed values of $k=1,2,3,...$ and $l=0,1,2,...,k-1$. Noticeably, the
positions for these subsequences can be asymptotically obtained from those
belonging to the superstable periodic orbits of lengths $F_{n}$ at $\Omega
_{n}$ or $\Omega _{n}^{\prime }$. In particular, the positions for the main
subsequence $k=1$, that constitutes the lower bound of the entire
trajectory, can be identified to be $\theta _{F_{n}}\simeq \left\vert
d_{n}\right\vert =$ $\left\vert \alpha _{gm}\right\vert ^{-n}$, $\Omega
=\Omega _{n}$. The positions for each time subsequence that leads a group, $%
k\geq 1$ and $l=0$, i.e. $\tau =kF_{n}$ in Figs. 3 can be expressed, when
use is made of the time shift $\tau =t+k$ and%
\begin{equation}
\frac{F_{n}}{F_{n-1}}\frac{F_{n-1}}{F_{n-2}}...\frac{F_{2}}{F_{1}}\simeq
w_{gm}^{-n},  \label{dressed2}
\end{equation}%
as 
\begin{equation}
\theta _{\tau }=\theta _{k}\exp _{Q_{+}}(\Lambda _{Q_{+}}^{(k)}t),
\label{q-exp1}
\end{equation}
with $Q_{+}=$ $1-\ln w_{gm}/\ln \left\vert \alpha _{gm}\right\vert $ and $%
\Lambda _{Q_{+}}^{(k)}=\ln \left\vert \alpha _{gm}\right\vert /k\ln w_{gm}$.
The same time subsequence positions given by $\tau =kF_{n}+1$ in Fig. 4 can
be expressed as%
\begin{equation}
\left\vert \Omega _{\infty }-\theta _{\tau }\right\vert =\left\vert \Omega
_{\infty }-\theta _{k}\right\vert \exp _{Q_{-}}(\Lambda _{Q_{-}}^{(k)}t),
\label{q-exp2}
\end{equation}%
with $Q_{-}=1-\ln w_{gm}/3\ln \left\vert \alpha _{gm}\right\vert $ and $%
\Lambda _{Q_{-}}^{(k)}=3\ln \left\vert \alpha _{gm}\right\vert /k\ln w_{gm}$%
. Interestingly, these results can be seen to satisfy the dynamical scaling
relation, 
\begin{equation}
h_{\tau }=\tau ^{1/(1-Q)}g(\tau ^{-1/(1-Q)}\theta )  \label{scaling1}
\end{equation}%
with $h_{\tau }=\theta _{\tau }$ or $\left\vert \Omega _{\infty }-\theta
_{\tau }\right\vert $, where $\tau =kw_{gm}^{-n}$ with $w_{gm}=$ $\alpha
_{gm}^{1-Q_{+}}$ or $w_{gm}=$ $\alpha _{gm}^{3(1-Q_{-})}$, and where $g$ is
the fixed-point map with $\theta =\theta _{k}$ or $\left\vert \Omega
_{\infty }-\theta _{k}\right\vert $. The similarities with the dynamical
properties of the supercycle orbit $2^{\infty }$ at the Feigenbaum attractor
are remarkable, see Refs. \cite{robledo1}, \cite{robledo2} for the quadratic
map and Refs. \cite{robledo3}, \cite{robledo8} for its generalization to
general nonlinearity $z>1$.

\begin{figure}[tbp]
\includegraphics[width=1.0\columnwidth,angle=-90,scale=0.8]{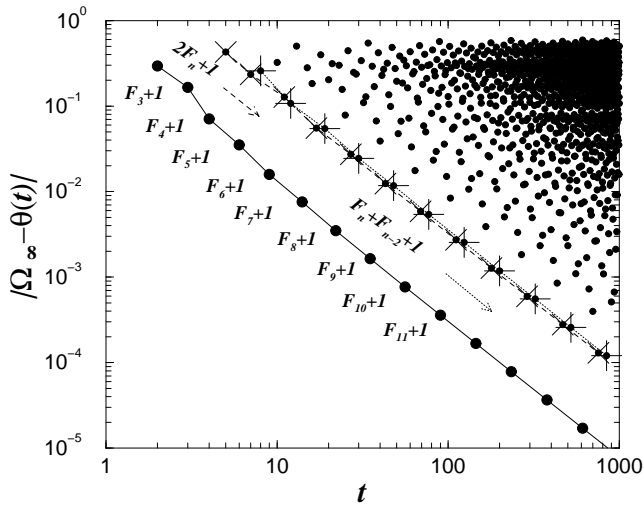} 
\caption{{\protect\small Positions }$\left\vert \Omega _{\infty }-\protect%
\theta (t)\right\vert ${\protect\small \ vs }$t${\protect\small \ in
logarithmic scales for the orbit with initial condition }$\protect\theta %
(0)=0${\protect\small \ at }$\Omega _{\infty }${\protect\small \ of the
critical circle map }$K=1${\protect\small . The labels indicate iteration
time }$t${\protect\small . The dashed line and symbols }$\times $%
{\protect\small \ correspond to times of the form }$2F_{n}+1${\protect\small %
, and the dotted line and symbols }$+${\protect\small \ correspond to times
of the form }$F_{n}+F_{n-2}+1$.{\protect\small \ The same results are
obtained with }$\Omega _{\infty }${\protect\small \ replaced by }$\Omega
_{\infty }^{\prime }${\protect\small . See text for description.} }
\end{figure}

\section{Sensitivity to initial conditions}

We have now elements to derive the sensitivity to initial conditions $\xi
_{t}(\theta _{0})$, defined as $\xi _{t}(\theta (0))\equiv d\theta
(t)/d\theta (0)$, when either $\theta (0)\simeq 0$ or $\theta (0)\simeq
\Omega _{\infty }$. For simplicity of presentation we consider the case of
the leading position subsequences within each group $k$, those for times $%
\tau =kF_{n}$ as described in the previous section. The results for the
other subsequences within each group are similar and we only quote the
differences in the expressions. We proceed in the following way. Recall \cite%
{fks1} that%
\begin{equation}
f_{n}(\theta )\simeq \alpha _{gm}^{-n}\ g(\alpha _{gm}^{n}\theta ),\;\theta
\simeq 0,  \label{fks1eq}
\end{equation}%
where $f_{n}(\theta )\equiv $ $f_{\Omega _{n},1}^{F_{n}}(\theta )-F_{n-1}$,
and consider the lower-order terms in the expansion of $g(\alpha
_{gm}^{n}\theta )$, i.e.%
\begin{equation}
g(\alpha _{gm}^{n}\theta )\simeq g(0)+\frac{1}{6}g^{\prime \prime \prime
}(0)\alpha _{gm}^{3n}\theta ^{3}.  \label{expg1}
\end{equation}%
If we denote the starting positions of two nearby trajectories at time $\tau
=k$ by 
\begin{equation}
\theta _{k}=f_{1}(\theta )\simeq g(0)+\frac{1}{6}g^{\prime \prime \prime
}(0)\theta ^{3}  \label{initial1}
\end{equation}%
and%
\begin{equation}
\phi _{k}=f_{1}(\phi )\simeq g(0)+\frac{1}{6}g^{\prime \prime \prime
}(0)\phi ^{3},  \label{initial2}
\end{equation}%
their positions at time $\tau =kF_{n}$ are given by%
\begin{equation}
\theta _{kF_{n}}=f_{n}(\theta )\simeq \alpha _{gm}^{-n}g(0)+\frac{1}{6}%
g^{\prime \prime \prime }(0)\alpha _{gm}^{2n}\theta ^{3}  \label{final1}
\end{equation}%
and%
\begin{equation}
\phi _{kF_{n}}=f_{n}(\phi )\simeq \alpha _{gm}^{-n}g(0)+\frac{1}{6}g^{\prime
\prime \prime }(0)\alpha _{gm}^{2n}\phi ^{3}.  \label{final2}
\end{equation}%
From the above we obtain 
\begin{equation}
\frac{\left\vert \theta _{kF_{n}}-\phi _{kF_{n}}\right\vert }{\left\vert
\theta _{k}-\phi _{k}\right\vert }=\ \left\vert \alpha _{gm}\right\vert
^{2n},\ n\gg 1,  \label{diff1}
\end{equation}%
where we have used the equality sign in consideration of the scaling limit
for large $n$. For each subsequence $k$, the sensitivity $\xi _{t}$, given
for the trajectories considered by 
\begin{equation}
\xi _{t}(\theta _{k})\equiv \lim_{\left\vert \theta _{k}-\phi
_{k}\right\vert \rightarrow 0}\frac{\left\vert \theta _{\tau }-\phi _{\tau
}\right\vert }{\left\vert \theta _{k}-\phi _{k}\right\vert },  \label{sens1}
\end{equation}%
can be written, with use of the shifted time variable $t\equiv \tau -k$, and
observing that 
\begin{equation}
\left\vert \alpha _{gm}\right\vert ^{2n}=\left( 1+\frac{t}{k}\right) ^{\frac{%
2\ln \left\vert \alpha _{gm}\right\vert }{\ln w_{gm}}},
\label{powerlawtrick1}
\end{equation}%
as the $q$-exponential 
\begin{equation}
\xi _{t}(\theta _{k})=\exp _{q}\left[ \lambda _{q}^{(k)}t\right] ,
\label{sens2}
\end{equation}%
where%
\begin{equation}
q=1+\frac{\ln w_{gm}}{2\ln \left\vert \alpha _{gm}\right\vert }\simeq
0.051003  \label{q-value1}
\end{equation}%
and 
\begin{equation}
\lambda _{q}^{(k)}=2\frac{\ln \left\vert \alpha _{gm}\right\vert }{k\ln
w_{gm}}.  \label{lambdaq1}
\end{equation}%
The value of \textit{q} in Eq. (\ref{q-value1}) agrees with that obtained in
earlier studies \cite{tsallis3}-\cite{lyra1}.

The departing positions, $\theta (0)\simeq 0$ (or $\theta (0)\simeq 1$), for 
$\xi _{t}$ in Eq. (\ref{sens2}) are located in the most crowded region of
the attractor and the value of $k$ in $\lambda _{q}^{(k)}$ in Eq. (\ref%
{sens2}) locates this position as the top position with time label $\tau
=kF_{1}=k$ of the subsequences described by the times $\tau =kF_{n}$ in
Figs. 3a and 3b. By construction, the observation times in Eq. (\ref{sens2})
are precisely $\tau =kF_{n}$ when the iterate is close to the most sparse
region of the attractor, $\theta _{0}\simeq \Omega _{\infty }$ (or $\theta
_{0}\simeq \Omega _{\infty }^{\prime }$). The sensitivity $\xi _{t+1}$ for
the inverse trajectories, with departing positions at the most sparse region
and with observation times when at the most crowded region, can be evaluated
in the same way as above, with the result%
\begin{equation}
\xi _{t+1}=\exp _{2-q}\left[ \lambda _{2-q}^{(k)}t\right] ,  \label{sens3}
\end{equation}%
with $q$ as above and where 
\begin{equation}
\lambda _{2-q}^{(k)}=-\frac{2\ln \left\vert \alpha _{gm}\right\vert }{%
kw_{gm}\ln w_{gm}}.  \label{lambdaq2}
\end{equation}%
Trajectories expand in one direction and contract in the opposite one. We
recall that $\exp _{q}(x)=1/\exp _{2-q}(-x)$. The factor of $w_{gm}^{-1}$ in 
$\lambda _{2-q}^{(k)}$ appears because of a basic difference between the
orbits of periods $F_{n}$ and $F_{\infty }$. In the latter case, to reach $%
\theta (t^{\prime })\simeq 0$ from $\theta (0)\simeq 1$ at times $t^{\prime
}=F_{n}+1$ the iterate necessarily moves into positions of the next period $%
F_{n+1}$, and orbit contraction is $w_{gm}^{-1}$ as effective than
expansion. When considering the other sets of positions within each group of
subsequences one obtains the same results as in Eqs. (\ref{sens2}) to Eq. (%
\ref{lambdaq2}) with $k$ replaced by $(k-l)+lw_{gm}^{2}$ in Eqs. (\ref%
{lambdaq1}) and (\ref{lambdaq2}).

\section{Mori's $q$-phase transitions and Tsallis' $q$ index}

About 15 years ago Mori and coworkers developed a broad thermodynamic
formalism to characterize sharp changes at bifurcations and at other
singular attractors in low dimensional maps \cite{mori1}. The formalism was
specially adapted to study critical attractors and was applied to the
specific case of the onset of chaos in the circle map at the golden mean 
\cite{mori3}, \cite{mori1} and in the logistic map at the period doubling
accumulation point \cite{mori2}, \cite{mori1}. For critical attactors the
scheme involves the evaluation of fluctuations of the generalized
finite-time Lyapunov coefficient 
\begin{equation}
\lambda (t,x_{0})=\frac{1}{\ln t}\sum_{i=0}^{t-1}\ln \left\vert \frac{%
df(x_{i})}{dx_{i}}\right\vert ,\quad t\gg 1,  \label{lyapunovdef}
\end{equation}%
where $f(x)$ is a map with control parameter(s) tuned at a critical
attractor. Notice the replacement of the customary $t$ by $\ln t$ in Eq. (%
\ref{lyapunovdef}), as the ordinary Lyapunov coefficient $\lambda _{1}$
vanishes for critical attractors at $t\rightarrow \infty $.

The density distribution for the values of $\lambda $, at $t\gg 1$, $%
P(\lambda ,t)$, is written in the form \cite{mori1}, \cite{mori3}, \cite%
{mori2} 
\begin{equation}
P(\lambda ,t)=t^{-\psi (\lambda )}P(0,t),
\end{equation}%
where $\psi (\lambda )$ is a concave spectrum of the fluctuations of $%
\lambda $ with minimum $\psi (0)=0$ and is obtained as the Legendre
transform of the `free energy' function $\phi (\mathsf{q})$, defined as $%
\phi (\mathsf{q})\equiv -\lim_{t\rightarrow \infty }\ln Z(t,\mathsf{q})/\ln
t $, where $Z(t,\mathsf{q})$ is the dynamic partition function 
\begin{equation}
Z(t,\mathsf{q})\equiv \int d\lambda \ P(\lambda ,t)\ t^{-(\mathsf{q}%
-1)\lambda }.  \label{partition1}
\end{equation}%
The `coarse-grained' function of generalized Lyapunov coefficients $\lambda (%
\mathsf{q})$ is given by $\lambda (\mathsf{q})\equiv d\phi (\mathsf{q})/d%
\mathsf{q}$ and the variance $v(\mathsf{q})$ of $P(\lambda ,t)$ by $v(%
\mathsf{q})\equiv d\lambda (\mathsf{q})/d\mathsf{q}$ \cite{mori1}, \cite%
{mori3}, \cite{mori2}. The functions $\phi (\mathsf{q})$ and $\psi (\lambda
) $ are the dynamic counterparts of the Renyi dimensions $D(\mathsf{q})$ and
the spectrum $f(\alpha )$ that characterize the geometric structure of the
attractor \cite{beck1}. The form $t^{-\psi (\lambda )}$ for $P(\lambda ,t)$
and the weight $t^{-(\mathsf{q}-1)\lambda }$ in Eq. (\ref{partition1}) are
intended for the description of the \textit{envelope} of the fluctuations at
the critical attactor \cite{mori1}, \cite{mori3}, \cite{mori2}.

As with ordinary thermal 1st order phase transitions, a `$q$-phase'\
transition is indicated by a section of linear slope $m_{c}=1-q$ in the
spectrum (free energy) $\psi (\lambda )$, a discontinuity at $\mathsf{q}=q$
in the Lyapunov function (order parameter) $\lambda (\mathsf{q})$, and a
divergence at $q$ in the variance (susceptibility) $v(\mathsf{q})$. For the
onset of chaos at golden mean a single $q$-phase transition was numerically
determined \cite{mori1} and found to occur approximately at a value around $%
m_{c}=-(1-q)\simeq -0.95$. It was pointed out in Ref. \cite{mori1} that this
value would actually be $m_{c}=-(1-q)=-\alpha _{\max }/2\simeq -0.948997$.
Our analysis below shows that this initial result gives a rough picture of
the dynamics at the golden mean attractor and that actually an infinite
family of $q$-phase transitions of decreasing magnitude takes place there.

From the results for $\lambda _{q}^{(k)}$ and $\lambda _{2-q}^{(k)}$ we can
construct the two-step Lyapunov function 
\begin{equation}
\lambda (\mathsf{q})=\left\{ 
\begin{array}{l}
\lambda _{q}^{(1)},\;-\infty <\mathsf{q}\leq q, \\ 
0,\;\;\;\;\;q<\mathsf{q}<2-q, \\ 
\lambda _{2-q}^{(1)},\;\;\;2-q\leq \mathsf{q}<\infty ,%
\end{array}%
\right.  \label{lambdaq3}
\end{equation}%
where $\lambda _{q}^{(1)}=2\ln \left\vert \alpha _{gm}\right\vert /\ln
w_{gm}\simeq 1.053744$ and $\lambda _{2-q}^{(1)}=-w_{gm}^{-1}\lambda
_{q}^{(1)}\simeq -1.704994$. See Fig. 5a. Direct contact can be established
now with the formalism developed by Mori and coworkers and the $q$-phase
transition reported in Ref. \cite{mori3}. The step function for $\lambda (%
\mathsf{q})$ can be integrated to obtain the spectrum $\phi (\mathsf{q})$ ($%
\lambda (\mathsf{q})\equiv d\phi /d\mathsf{q}$) and from this its Legendre
transform $\psi (\lambda )$ ($\equiv \phi -(1-\mathsf{q})\lambda $). The
free energy functions $\phi (\mathsf{q})$ and $\psi (\lambda )$ obtained
from the two-step $\lambda (\mathsf{q})$ determined above are given by 
\begin{equation}
\phi (\mathsf{q})=\left\{ 
\begin{array}{l}
\lambda _{q}^{(1)}(\mathsf{q}-q),\;\;\mathsf{q}\leq q, \\ 
0,\;\;\;\;\;\;q<\mathsf{q}<2-q, \\ 
\lambda _{2-q}^{(1)}(\mathsf{q}-2+q),\;\mathsf{q}\geq 2-q,%
\end{array}%
\right.  \label{phiq3}
\end{equation}%
and 
\begin{equation}
\psi (\lambda )=\left\{ 
\begin{array}{l}
-(1-q)\lambda ,\;\lambda _{2-q}^{(1)}<\lambda <0, \\ 
(1-q)\lambda ,\;0<\lambda <\lambda _{q}^{(1)}.%
\end{array}%
\right.  \label{psiq3}
\end{equation}%
See Fig. 5b. The constant slopes of $\psi (\lambda )$ represent the $q$%
-phase transitions associated with trajectories linking two regions of the
attractor, $\theta \simeq 0$ (or $\theta \simeq 1)$ and $\theta \simeq
\Omega _{\infty }$, and their values $1-q$ and $q-1$ correspond to the index 
$q$ obtained for the $q$-exponentials $\xi _{t}$ in Eqs. (\ref{sens2}) and (%
\ref{sens3}). The slope $q-1\simeq -\alpha _{\max }/2\simeq -0.949$
coincides with that originally detected in Ref. \cite{mori3}.

\begin{figure}[tbp]
\includegraphics[width=.9\columnwidth,angle=-90,scale=0.7]{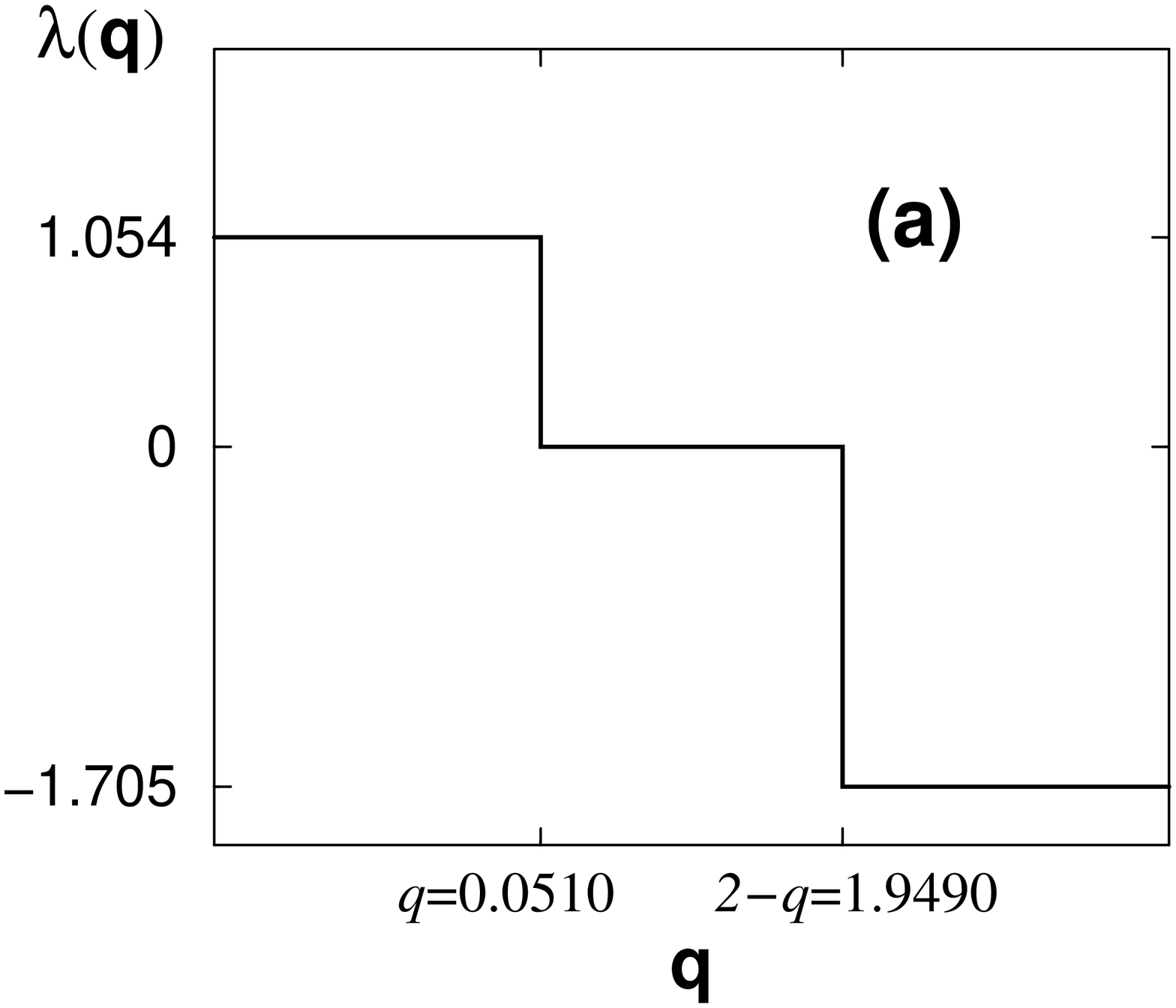} %
\includegraphics[width=.9\columnwidth,angle=-90,scale=0.7]{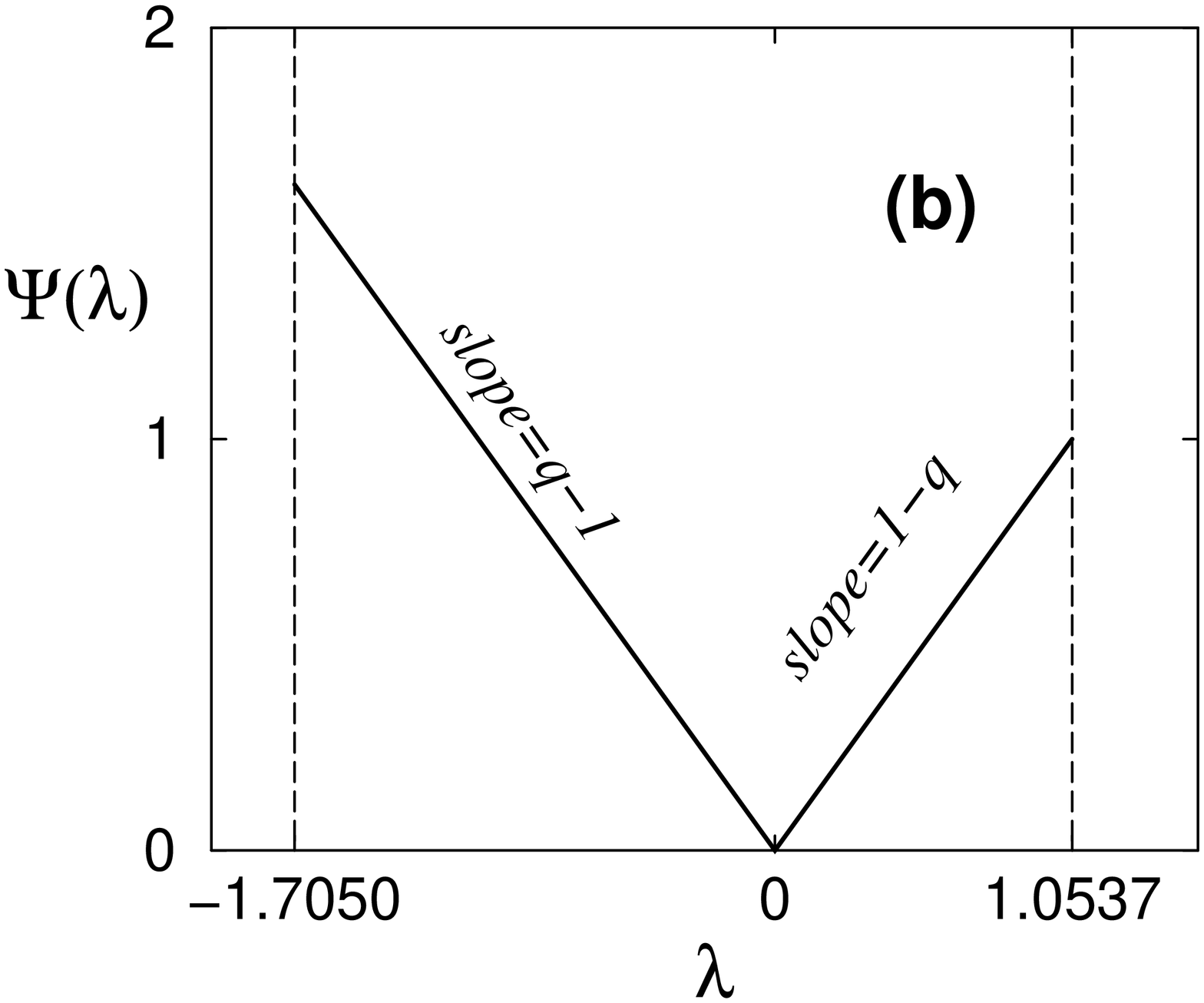} 
\caption{$q${\protect\small -phase transitions occurring at index values }$q$%
{\protect\small \ and }$2-q${\protect\small . (a) The two-step Lyapunov
function }$\protect\lambda (\mathsf{q})${\protect\small . (b) The piece-wise
spectrum }$\protect\psi (\protect\lambda )${\protect\small . These
transitions correspond to the main discontinuity in the trajectory scaling
function }$\protect\sigma (y)${\protect\small . See text for details.} }
\end{figure}

\section{Multifractal dynamics from the trajectory scaling function}

The trajectory scaling function $\sigma (y)$ for the golden mean route is
constructed \cite{feigenbaum2}, \cite{mainieri1} via the evaluation of the
ratios 
\begin{equation}
\sigma _{n}(m)=\frac{d_{n+1,m}}{d_{n,m}},  \label{sigma1}
\end{equation}%
where $d_{n,m}=\left\vert \theta _{m}-\theta _{F_{n}+m}\right\vert $, $%
m=0,1,2...$, and where the denominator is replaced by $d_{n,m}=\left\vert
\theta _{m}-\theta _{m-F_{n}}\right\vert $ in case $m\geq F_{n}$. The
function $\sigma (y)$ is obtained as the limit $\sigma (y)\equiv
\lim_{n\rightarrow \infty }$ $\sigma _{n}(m)$ where $y=\lim_{n\rightarrow
\infty }m/F_{n}$, and shows finite (jump) discontinuities at all rationals
of the form $m/F_{n}$. The main discontinuity appears at $y=0$, since $%
\sigma (0)=\left\vert \alpha _{gm}\right\vert ^{-1}$ but $\sigma
(0^{+})=\left\vert \alpha _{gm}\right\vert ^{-3}$, and again at $%
y=w_{gm}^{2} $. Other discontinuities in $\sigma (y)$ appear at $y=w_{gm}$, $%
w_{gm}^{3}$, $w_{gm}^{4}$, $w_{gm}+w_{gm}^{3}$, $w_{gm}+w_{gm}^{4}$,$\ $%
etc., and of course $w_{gm}+w_{gm}^{2}=1$. We leave a detailed account of
these properties to a later occasion \cite{robledo9}. In most cases it is
only necessary to consider the first few as their magnitude decreases
rapidly. See, e.g. Fig. 2 in Ref. \cite{mainieri1}. The discontinuities of $%
\sigma _{n}(m)$ can be suitably obtained by first generating the superstable
orbit at $\Omega _{\infty }$ and then plotting the position differences $%
\left\vert \theta _{\tau }-\theta _{m}\right\vert $ for times of the form $%
\tau =F_{n}+m$, $n=0,1,2,...$, in logarithmic scales. The distances that
separate positions along the time subsequence correspond to the logarithm of
the distances $d_{n,m}$. See Figs. 3 and 4 where the constant spacing of
positions along the main diagonal provide the values for $\ln d_{n,0}\simeq
-n\ln \left\vert \alpha _{gm}\right\vert $ and $\ln d_{n,1}\simeq -3n\ln
\left\vert \alpha _{gm}\right\vert $, respectively. The constant slope $%
s_{m} $ of the resulting time subsequence data is related to $\sigma _{n}(m)$%
, i.e. $\sigma _{n}(m)=(w_{gm})^{s_{m}}$. See Figs. 3 and 4 where the slopes
of the main diagonal subsequences have the values $-\ln \left\vert \alpha
_{gm}\right\vert /\ln w_{gm}$ and $-3\ln \left\vert \alpha _{gm}\right\vert
/\ln w_{gm}$, respectively. From these two slopes the value of the largest
jump discontinuity of $\sigma _{n}(m)$ is determined.

The sensitivity $\xi _{t}(\theta (0))$ can be evaluated for trajectories
within the attractor via consideration of the discontinuities of $\sigma (y)$%
. We fix the initial and the final separation of the trajectories to be the
diameters $\Delta \theta (0)=d_{n,m}$ and $\Delta \theta (t)=d_{n,m+t}$, $%
t=F_{n}-1$, respectively. Then, $\xi _{t}(\theta (0))$ is obtained as 
\begin{equation}
\xi _{t}(\theta (0))=\lim_{n\rightarrow \infty }\left\vert \frac{d_{n,m+t}}{%
d_{n,m}}\right\vert .
\end{equation}%
Notice that in this limit $\Delta \theta (0)\rightarrow 0$, $t\rightarrow
\infty $ \textit{and} the $F_{n}$-supercycle becomes the onset of chaos.
Then, $\xi _{t}(\theta (0))$ can be written as 
\begin{equation}
\xi _{t}(m)\simeq \left\vert \frac{\sigma _{n}(m-1)}{\sigma _{n}(m)}%
\right\vert ^{n},\ t=F_{n}-1\;\text{and}\;n\;\text{large},  \label{sens4}
\end{equation}%
where we have used $\left\vert \sigma _{n}(m)\right\vert ^{n}\simeq
\prod_{i=1}^{n}\left\vert d_{i+1,m}/d_{i,m}\right\vert $ and $%
d_{i+1,m+F_{n}}=d_{i+1,m}$. Notice that for the inverse process, starting at 
$\Delta \theta (0)=d_{n,m+t}=d_{n,m-1}$ and ending at $\Delta \theta
(t^{\prime })=d_{n,m}=d_{n,m-1+t^{\prime }}$, with $t^{\prime }=F_{n}+1$ one
obtains 
\begin{equation}
\xi _{t^{\prime }}(m-1)\simeq \left\vert \frac{\sigma _{n}(m)}{\sigma
_{n}(m-1)}\right\vert ^{n},\ t^{\prime }=F_{n}+1\;\text{and}\;n\;\text{large}%
.  \label{sens5}
\end{equation}

Therefore, from the properties of $\sigma (y)$ we can determine those for $%
\xi _{t}$. This function can be obtained at different levels of
approximation by considering only a fixed number of its larger
discontinuities. The discontinuities of $\sigma (y)$ appear divided into two
sets, the first falls on the interval $0\leq y<w_{gm}$, and the second on $%
w_{gm}\leq y<1$. The amplitudes of the jumps in the second set are much
smaller than those in the first set \cite{mainieri1}. The simplest
approximation for $\sigma (y)$ is to consider only one jump in the first set
at $w_{gm}^{2}$ and none in the second set, so that%
\begin{equation}
\sigma (y)=\left\{ 
\begin{array}{c}
\alpha _{0}^{-1},\;0\leq y<w_{gm}^{2}, \\ 
\alpha _{1}^{-1},\;w_{gm}^{2}\leq y<1%
\end{array}%
\right. ,  \label{sigma2}
\end{equation}%
where $\alpha _{0}=\left\vert \alpha _{gm}\right\vert ^{3}\simeq 2.139583$
and $\alpha _{1}=\left\vert \alpha _{gm}\right\vert \simeq 1.288575$. This
level of approximation assumes that there are only two scaling factors,
those corresponding to the most crowded and most sparse regions of the
attractor. The next level of approximation for $\sigma (y)$ involves two
jumps in the first set at $w_{gm}^{3}$ and $w_{gm}^{2}$, and still none in
the second set, so that%
\begin{equation}
\sigma (y)=\left\{ 
\begin{array}{c}
\alpha _{0}^{-1},\;0\leq y<w_{gm}^{3}, \\ 
\alpha _{1}^{-1},\;w_{gm}^{3}\leq y<w_{gm}^{2}, \\ 
\alpha _{2}^{-1},\;w_{gm}^{2}\leq y<w_{gm} \\ 
\alpha _{3}^{-1},\;w_{gm}\leq y<1%
\end{array}%
\right. ,  \label{sigma3}
\end{equation}%
where $\alpha _{0}=\left\vert \alpha _{gm}\right\vert ^{3}$, $\alpha
_{1}=1/\sigma (w_{gm}^{2})\simeq 2.452$, $\alpha _{2}=1/\sigma
(w_{gm})\simeq 1.430$ and $\alpha _{3}=\left\vert \alpha _{gm}\right\vert $ 
\cite{mainieri1}. The two additional scaling factors correspond to two
`midway' regions between the most crowded and most sparse regions of the
attractor. More accurate approximations to $\sigma (y)$ can be constructed
by incorporation of additional discontinuities. Note that the number of
discontinuities increase in each set as $F_{M}-1$, $M=1,2,..$.

Use of Eq. (\ref{sigma2}) in Eqs. (\ref{sens4}) and (\ref{sens5}) (together
with Eq. (\ref{powerlawtrick1})) recovers our previous results for $\xi _{t}$
and $\lambda _{q}$ in Eqs. (\ref{sens2}) to (\ref{lambdaq2}), and therefore
also those for\ $\lambda (\mathsf{q})$, $\phi (\mathsf{q})$ and $\psi
(\lambda )$ in Eqs. (\ref{lambdaq3}) to (\ref{psiq3}). When use is made of
Eq. (\ref{sigma3}) in Eqs. (\ref{sens4}) and (\ref{sens5}) we obtain three
values for the $q$ index, $q_{0}$, $q_{1}$ and $q_{2}$ (together with the
conjugate values $2-q_{0}$, $2-q_{1}$ and $2-q_{2}$ for the inverse
trajectories). For each value of $q$ there is a set of $q$-Lyapunov
coefficients running from a maximum $\lambda _{q_{j},\max }$ to zero (or a
minimum $\lambda _{2-q_{j},\min }$ to zero). From the results for $\lambda
_{q_{j}}^{(k)}$ and $\lambda _{2-q_{j}}^{(k)}$, $j=0,1,2$, we can construct
three Lyapunov functions $\lambda _{j}(\mathsf{q})$, $-\infty <\mathsf{q}%
<\infty $, each with two jumps located at $\mathsf{q}=q_{j}=1-\ln 2/\ln
\alpha _{j}/\alpha _{j+1}$ and $\mathsf{q}=2-q_{j}$. We obtain a couple of
two $q$-phase transitions for each of the three values of the $q$ index, $%
q_{0}$, $q_{1}$ and $q_{2}$. The constant slope values for the $q$-phase
transitions at $1-q_{0}$ and $q_{0}-1$ appear again, but now we have two
other pairs of transitions with slope values $1-q_{1}$ and $q_{1}-1$, and, $%
1-q_{2}$ and $q_{2}-1$, that correspond, respectively, to orbits that link
the most crowded region of the attractor to the `medium crowded' region, and
to orbits that link the `medium sparse' region with the `most sparse' region
of the attractor. Similar but more elaborated results are obtained when more
discontinuities in $\sigma (y)$ are taken into account. These results
reflect the multi-region nature of the multifractal attractor and the memory
retention of these regions in the dynamics. Thus, when $\sigma (y)=\sigma
(y+)$ one has $\xi _{t}=1$ (or $\lambda _{q}(\theta (0))=0$) which
corresponds to trajectories that depart and arrive in the same region, when $%
\sigma (y)\neq \sigma (y+)$ the power laws in Eqs. (\ref{sens4}) and (\ref%
{sens5}) correspond to a departing position in one region and arrival at a
different region and vice versa, the trajectories expand in one sense and
contract in the other.

\section{Summary and assessment of applicability of $q$-statistics}

We have examined in detail the structure of the superstable orbit at the
golden mean critical attractor for zero-slope cubic inflection point maps.
The knowledge of this structure allowed us to determine the sensitivity to
initial conditions for sets of starting positions within the attractor. We
found that $\xi _{t}$ is made up of a hierarchy of families of infinitely
many interconnected $q$-exponentials. Each pair of regions in the
multifractal attractor, that contain the starting and finishing positions of
a set of trajectories, leads to a family of $q$-exponentials with a fixed
value of the index $q$ and an associated spectrum of $q$-Lyapunov
coefficients $\lambda _{q}^{(k)}$. The indexes $q$ and the spectra $\lambda
_{q}^{(k)}$ come in pairs, $q$ and $2-q$, and $\lambda _{q}^{(k)}$ and $%
\lambda _{2-q}^{(k)}$, as these correspond to switching starting and
finishing trajectory positions. This dynamical organization is difficult to
resolve from the consideration of a straightforward time evolution, i.e. the
record of positions at every time $t$ for a trajectory started at an
arbitrary position $\theta (0)$ within the attractor. In this case what is
observed \cite{mori3} are strongly fluctuating quantities that persist in
time with a scrambled pattern structure that exhibits memory retention.
Unsystematic averages over $\theta (0)$ would rub out the details of the
multiscale dynamical properties we uncovered. On the other hand, if specific
initial positions with known location within the multifractal are chosen,
and \ subsequent positions are observed only at pre-selected times, when the
trajectories visit another selected region, a distinct $q$-exponential
expression for $\xi _{t}$ is obtained.

Subsequently, we explained that the above described properties for the
sensitivity relate to sets of dynamical phase transitions. That is, the
dynamics on the attractor consists of an infinite family of Mori's $q$-phase
transitions, each associated, as before, to trajectories that have common
starting and finishing positions located at specific regions of the
attractor. The specific values of the variable $\mathsf{q}$ in the formalism
of Mori and colleagues at which the $q$-phase transitions take place are the
same values for the Tsallis entropic index $q$ in $\xi _{t}$. The
transitions come in pairs at $q$ and $2-q$ as they are tied down to the
expressions for $\lambda _{q}^{(k)}$ in $\xi _{t}$. The dominant dynamical
transition is associated with movement from the most crowded to the most
sparse regions of the attractor (or vice versa) and the values of its
relevant parameters coincide with those found in an earlier study with the
use of Mori's formalism \cite{mori3}. Here we have determined the $q$-phase
transitions via the direct evaluation of the spectra $\lambda _{q}^{(k)}$
and $\lambda _{2-q}^{(k)}$ in $\xi _{t}$. Also, we established a connection
between the trajectory scaling function $\sigma $ and the expressions for $%
\xi _{t}$, such that the discontinuities of the former determine the $q$%
-phase transitions and the values for the $q$ index. Conversely, from the $q$%
-exponential expressions for $\xi _{t}$ the scaling function $\sigma $ can
be conveniently determined.

Our results clearly apply to many other families of maps with zero slope
inflection points of cubic nonlinearity and are likely to hold also for
general nonlinearity $z>1$ \cite{delbourgo1}. These results may be of
relevance in condensed matter problems where the quasiperiodic route to
chaos is involved. One interesting example is the phenomenon of the
localization transition for transport in incommensurate systems, where Schr%
\"{o}dinger equations with quasiperiodic potentials \cite{harper1} are
equivalent to nonlinear maps with a quasiperiodic route to chaos, and where
the divergence of the localization length translates into the vanishing of
the ordinary Lyapunov coefficient \cite{satija1}. The dynamics at this type
of multifractal critical attractor is also represented by a $\xi _{t}$ given
by families of interconnected $q$-exponentials with $q$-generalized spectra
of Lyapunov coefficients $\lambda _{q}^{(k)}$. The latter quantities yield
information on the power-law decay rates of multifractal-structured wave
functions at the localization transition. Our findings reported here are
very similar to those recently determined for the Feigenbaum attractor in
unimodal maps \cite{robledo8}, and reinforce the notion of generality of our
identification of the source for the entropic index $q$ observed at critical
attractors.

\subsection{The utility of $q$-statistics}

Before concluding our analysis it is of interest to reconsider our
developments in such a way as to illustrate the incidence and usefulness of
the $q$-statistics in the description of the dynamics on the attractor at
the transition to chaos. We return to the dynamical partition function in
Eq. (\ref{partition1}) and rewrite it as%
\begin{equation}
Z(\tau ,\mathsf{q})\equiv \int d\lambda \ P(\lambda ,\tau )\ W(\lambda ,\tau
)^{1-\mathsf{q}},  \label{partition2}
\end{equation}%
where the weight $W(\lambda ,\tau )$ was assumed by Mori and colleagues \cite%
{mori1} to have the form $W(\lambda ,\tau )=\exp (\lambda \tau )$ for
chaotic attractors and $W(\lambda ,\tau )=\tau ^{\lambda }$ for the envelope
of fluctuations of the attractor at the onset of chaos, with $\lambda $ a
generalized Lyapunov coefficient. The variable $\mathsf{q}$ plays the role
of a 'thermodynamic field' like the external magnetic field in a thermal
magnet or, alternatively, $1-\mathsf{q}$ can be thought analogous to the
inverse temperature. We note that $W(\lambda ,\tau )$ can be written in both
cases as%
\begin{equation}
W(\lambda ,\tau )=\left\vert \frac{df^{(\tau )}(\theta _{0})}{d\theta _{0}}%
\right\vert   \label{weight1}
\end{equation}%
since $\lambda (\tau ,\theta _{0})=A(\tau )\ln \left\vert df^{(\tau
)}(\theta _{0})/d\theta _{0}\right\vert $ with $A(\tau )=$ $\tau ^{-1}$ for
chaotic attractors and $A(\tau )=(\ln \tau )^{-1}$ for the envelope of the
fluctuating sensitivity of the critical attractor \cite{mori1}. For chaotic
attractors the fluctuations of $\lambda (\tau ,\theta _{0})$ die out as $%
\tau \rightarrow \infty $ and this quantity becomes the ordinary Lyapunov
coefficient $\lambda _{\infty }$ independent of the initial position $\theta
_{0}$. For critical attactors $\lambda (\tau ,\theta _{0})$ maintains its
dependence on $\tau $ and $\theta _{0}$ but as we have seen $\lambda (\tau
,\theta _{0})$ is a well-defined constant \textit{provided} $\tau $ takes
values from a time subsequence $\tau (n,k,l)$, as explained in Sections 3
and 4 and shown in Figs. 3 and 4. For simplicity we consider now only the
case $l=0$, i.e. the subsequences $\tau (n,k)=$ $kF_{n}$, $n=1,2,3,...$ for
fixed values of $k=1,2,3,...$, but it is important to note that every
natural number $\tau $ appears in the complete set of subsequences. One
branch of the envelope of $\xi _{t}$ with $\theta _{0}=0$ is visited at
times $\tau $ of the form $\tau =$ $F_{n}$ for which we find $\lambda
(F_{n},0)=$ $\lambda _{q}^{(1)}$ while for other time subsequences one has $%
\lambda (kF_{n},0)=$ $\lambda _{q}^{(k)}$ (see Eq. (\ref{lambdaq1}). Again
for simplicity we ignore the inverse trajectories that lead to the conjugate 
$\lambda _{2-q}^{(k)}$ (see Eq. (\ref{lambdaq2})).

As we have proved, along each time subsequence the sensitivity $\xi _{t}=$ $%
\left\vert df^{(\tau )}(\theta _{0})/d\theta _{0}\right\vert $ is given by a 
$q$-exponential, like in Eq. (\ref{sens1}), and consequently we can express $%
\lambda (\tau ,\theta _{0})$ in terms of the $q$-deformed logarithm $\lambda
(\tau ,\theta _{0})=\tau ^{-1}\ln _{q}\left\vert df^{(\tau )}(\theta
_{0})/d\theta _{0}\right\vert $ with $q$ given by e.g. Eq. (\ref{q-value1}).
Therefore for times of the form $\tau (n,k)=$ $kF_{n}$ we have%
\begin{equation}
P(\lambda ,\tau )W(\lambda ,t)=\delta (\lambda -\lambda _{q}^{(k)})\exp
_{q}(\lambda _{q}^{(k)}t),  \label{weight2}
\end{equation}%
with $t=\tau -k$, and 
\begin{equation}
Z(t,\mathsf{q})=W(\lambda _{q}^{(k)},t)^{1-\mathsf{q}}=\left[ 1+(1-q)\lambda
_{q}^{(k)}t\right] ^{(1-\mathsf{q})/(1-q)}.  \label{partition3}
\end{equation}%
Notice that in Eq. (\ref{partition3}) $\mathsf{q}$ is a running variable
while $q$ is fixed as in Eq. (\ref{q-value1}).

The next step is to consider the uniform probability distribution, for fixed 
$\lambda _{q}^{(k)}$ and $t$, given by $p_{i}(\lambda _{q}^{(k)},t)=%
\overline{W}^{-1}$, $i=1,...,\overline{W}$, where $\overline{W}$ is the
integer nearest to (a large) $W$. A trajectory starting from $\theta _{0}=0$
visits a new site $\theta _{\tau }$ belonging to the time subsequence $\tau =
$ $kF_{n}$ every time the variable $n$ increases by one unit and never
repeats one. Each time this event occurs phase space is progressively
covered with an interval $\Delta \theta _{n}=\theta _{kF_{n+1}}-\theta
_{kF_{n}}$. The difference of the logarithms of the times between any two
such consecutive events is the constant $\ln w_{gm}^{-1}$, $n$ large,
whereas difference of the logarithms of the corresponding phase space
distances $\theta _{kF_{n+1}}$ and $\theta _{kF_{n}}$ is the constant $2\ln
\left\vert \alpha _{gm}\right\vert $. We define $W_{n}$ to be the total
phase space distance covered by these events up to time $\tau =$ $kF_{n}$.
In logarithmic scales this is a linear growth process in space and time from
which the constant probability $\overline{W}_{n}^{-1}$ of the uniform
distribution $p_{i}$ is defined. In ordinary phase space $\theta $ and time $%
t$ we obtain instead $W^{-1}=\exp (-\lambda _{q}^{(k)}t)$. We then have that 
\begin{equation}
Z(t,\mathsf{q})=\sum_{i=1}^{\overline{W}}p_{i}^{\mathsf{q}}=1+(1-\mathsf{q}%
)S_{\mathsf{q}},  \label{partition4}
\end{equation}%
where $S_{\mathsf{q}}=\ln _{\mathsf{q}}\overline{W}$ is the Tsallis entropy
for $p_{i}$.

The usefulness of the $q$-statistical approach is now evident when we recall
from our discussion in the previous section that the dynamics \textit{on}
the critical attractor is constituted (in its entirety) by a discrete set of 
$q$-phase transitions and that \textit{at} each of them the field variable $%
\mathsf{q}$ takes a specific value $\mathsf{q}=q$. At each such transition
both $Z$ and $S_{q}$ grow linearly with time along the subsequences $\tau
(n,k)$. In addition, $\lambda _{q}^{(k)}$ can be determined from $%
S_{q}/t=\lambda _{q}^{(k)}$. Thus, with the knowledge we have gained, a
convenient procedure for determining the fluctuating dynamics at a critical
multifractal attractor could take advantage of the above properties.

\textbf{Acknowledgments}. Partially supported by CONACyT and DGAPA-UNAM,
Mexican agencies.

\end{document}